\documentclass[12pt,preprint,deluxe]{aastex}

\usepackage{epsfig}

\begin{document}
\voffset-1cm

\newcommand{\gsim}{\hbox{\rlap{$^>$}$_\sim$}}
\newcommand{\lsim}{\hbox{\rlap{$^<$}$_\sim$}}

\title{The GRB/XRF-SN Association}
\author{Arnon Dar\altaffilmark{1}}

\altaffiltext{1}{arnon@physics.technion.ac.il\\
Physics Department and Space Research Institute, Technion, Haifa 32000,
Israel and}

% \maketitle

\begin{abstract}

There is mounting evidence that long duration gamma ray bursts (GRBs) and
X-ray flashes (XRFs) are produced by highly relativistic and narrowly
collimated jets ejected in core collapse supernova (SN) explosions akin to
SN~1998bw. We review the history of the GRB-SN association idea and its
observational verification. We summarize the present evidence for a
GRB/XRF-SN association. We comment on the possibility that most, perhaps all,
SN explosions produce GRBs, including SNe of Type Ia which may produce
short GRBs/XRFs. We list the major open questions that follow from a
GRB/XRF-SN association. Possible uses of the GRB-SN association in
cosmology are pointed out.

\end{abstract}


\section{History of the GRB/XRF-SN association}

The history of scientific breakthroughs is as fascinating as the 
breakthroughs themselves. The history of the GRB/XRF-SN association is 
not an exception.   

The production of gamma rays in supernova (SN) explosions was 
suspected (Colgate 1959 unpublished) long before the discovery of gamma
ray bursts (GRBs) in 1967 by the Vela satellites.  Klebesadel, Strong \&
Olson (1973), in their discovery paper, reported on a catalog search of SNe
coincident in time and sky position with the first 13 detected GRBs.
Colgate's suggestion (1968, 1974) that the breakout of the shock wave from
the stellar surface in core-collapse SNe may produce GRBs could be true,
if GRBs were observable only from relatively small cosmological distances.
This is because there are more than $10^6$ SN explosions per day in the
observable universe and only 2-3 GRBs per day. However, the Solar Maximum
Mission Satellite (SMM) did not detect a prompt GRB from the nearest SN
explosion in recent times, SN~1987A in the Large Magellanic Cloud, which
happened to be in its field of view (Chupp et al.~1987).  Dar \& Dado
(1987) considered the possibility that radiative decay of neutrinos from
core-collapse SN explosions would produce GRBs and used the disparity
between the cosmic SN and GRB rates, as well as SN~1987A, to derive bounds on
$\nu$ radiative decay. Goodman, Dar \& Nussinov, who studied the
production of $e^+e^-$ pairs by $\nu\,\bar\nu$ annihilation above the
neutrino sphere in core collapse supernova explosions, estimated that
neutrino annihilation in accretion induced collapse of neutron stars (ns),
or in ns-ns mergers in close binaries due to gravitational wave emission,
could form relativistic fireballs (Paczynski 1986; Goodman 1986) that
produce cosmological GRBs\footnote{The idea that neutrino annihilation in
ns-ns mergers produces cosmological GRBs was reproduced later by Eichler et
al. 1989.}. Yet, the authors pointed out that baryon contamination of the
fireball (now known as ``baryon-load'') poses a severe problem for this
mechanism.

The first strong indication that GRBs are indeed cosmological in origin
came from their isotropic sky distribution and their peak intensity
distribution measured by BATSE on board the Compton Gamma Ray Burst
Observatory (CGRO) shortly after its launch (Meegan et al. 1992). 
Following these observations, cosmological fireballs produced by neutrino
annihilation in ns-ns mergers became the leading candidate for the origin
of cosmological GRBs (see, e.g., Fishman \& Meegan~1995 and references 
therein). Dar et al.~(1992), however, suggested that 
neutrino-annihilation around ``compact
supernova'' produces cosmological GRBs, but Woosley (1993a,b) argued that
cosmological GRBs {\it cannot} be produced in SN explosions, and proposed
instead that they are produced by mildly relativistic jets ejected in the
direct collapse of massive stars into black holes {\it without} an associated
supernova, which he dubbed {\it failed supernovae} or {\it collapsars}
\footnote{This GRB-SN {\it dissociation} has been referenced in many recent
articles as the original suggestion of ... a GRB-SN {\it association}.}.
Finally, Shaviv and Dar (1995) suggested that highly relativistic jets
ejected in core collapse supernovae, or in supernova explosions driven by
accretion/merger induced collapse of compact stars in close binaries, can
produce the cosmological GRBs by inverse Compton scattering of circumburst
light. This idea was later incorporated into the cannonball (CB) model 
of GRBs (Dar \& De R\'ujula 2000; 2003).

The approximate localization of GRBs by the Beppo-SAX satellite (Costa et
al.~1997) led to discovery that long duration GRBs have afterglows (AGs),
i.e., long term emission of radiation at longer wavelengths\footnote{
Detection of a $\gamma$-ray emission at GeV and sub-GeV energies with
EGRET on board CGRO, during the burst and  up to $\sim2$h after it, has
been reported before by Hurley (1994).} (X-Rays: Costa et al.~1997;
Optical band: Van Paradijs et al.~1997; Radio band: Kulkarni et al.~1998).
These afterglows led to the precise sky localization of GRBs, the
discovery of their host galaxies and the measurements of their redshifts.
Long GRBs were found to be located in star formation regions of normal
star forming galaxies at large cosmological distances.  The approximate
power-law decline with time of their afterglows has been used to conclude
(e.g., Wijers et al.~1997; Piran 1999) that GRBs are produced in
relativistic fireballs and their afterglows are synchrotron radiation from
shock accelerated electrons in the collision of the fireball (or its
accelerated ejecta) with the interstellar medium (Paczynski \& Rhoads
1993; Katz 1994a,b; Meszaros \& Rees 1997). Arguments that the
observational data actually indicate that GRBs and their afterglows are
produced by narrow relativistic jets similar to those observed in quasars
and microquasars (Dar 1997; 1998) rather than by relativistic fireballs
were initially ignored or dismissed by the majority of the GRB community.

The localization of long duration GRBs in star formation regions, which
more than 90\% of the SNe take place, and not in the galactic halos where
most binary neutron stars are expected to reach long before they merge by
gravitational wave emission, was the first observational indication that
long duration GRBs may be produced by the death of massive stars in SN
explosions. The first direct evidence for an SN-GRB association came
a little later from the discovery by Galama et al.~(1998) of the very bright
SN~1998bw, at redshift $z=0.0085$, within the Beppo-SAX error circle
around the measured position of GRB~980425 (Soffita et al.~1998; Pian et
al.~1999). Its light curve indicated that the time of the SN explosion was
within $-2$ to $+0.7$ days of the GRB (Galama et al.~1998, Iwamoto et
al.~1998). This evidence did not fit at all into the framework of the
fireball model (FB) of GRBs.  The total equivalent isotropic $\gamma$-ray
energy release, $\sim 8\times 10^{47}$ erg, was some 5 orders of magnitude
smaller than that expected from a ``classical" GRB at $z=0.0085$. The
ruling majority concluded that either SN~1998bw and GRB~980425 were not
physically connected or that, if they were, they represent a new subclass
of rare events (e.g.~Bloom et al.~1998; Woosley, Eastman \& Schmidt~1999; 
Hurley et al.~2002; Galama 2003). These would be associated with what
Paczynski (1998a) and Iwamoto et al.~(1998) called ``hypernovae":
super-energetic explosions with kinetic energy exceeding $10^{52}$ erg, as
was inferred for SN~1998bw from its high expansion velocity and luminosity
(Patat et al.~2001), and from the very strong radio emission from its
direction (Kulkarni et al.~1998). 

Yet, a physical association between GRB~980425 and SN~1998bw was consistent
with GRB~980425 being a normal GRB produced by a highly relativistic jet
ejected in SN explosion and viewed off axis ((Shaviv \& Dar~1995; Wang \&
Wheeler~1998) at a viewing angle which is a few times larger than the
typical viewing angle of ordinary GRBs. Likewise, core collapse SN can be
highly asymmetric and SN~1998bw could appear to be an unusually bright SN,
perhaps because it was viewed near axis (Hoeflich, Wheeler \& Wang~1999).
These possibilities were first ignored by the majority of the GRB and SN
communities, which were accustomed to spherical models of SN explosions
and GRB ``fireballs''.

SN~1998bw was initially classified as Type Ib (Sadler et al.~1998) and
later as a peculiar Type Ic (Filippenko 1998; Patat and Piemonte 1998,
Patat et al.~2001). Its discovery initiated intensive searches of
positional and approximate temporal coincidences between GRBs and SN
explosions (e.g.~Kippen et al.~1998), in particular Type Ib and Ic SNe
(Woosley, Eastman \& Schmidt 1999). The search yielded two inconclusive
associations, one of them, between the peculiar Type II SN~1997cy, at
$z=0.063$, and the short-duration ($\sim 0.2$ s) GRB~970514 (Germany et
al.~2000) and another one between the Type Ic SN~1999E at $z=0.0261$ and
the long-duration GRB~980910 (Thorsett \& Hogg~1999; Rigon et al.~2003). 
Nevertheless, Dar \& Plaga(1999) \footnote{That paper was submitted to
Nature in summer 1998}) and Dar \& De R\'ujula~(1999) advocated the view
that most core-collapse SN explosions result in GRBs. 

Core collapse SNe are far from being standard candles. In particular SNe
of type Ib/Ic have light curves, peak intensity and peak time which
display a large dispersion. But if they are axially as opposed to
spherically symmetric ---as they would be if a fair fraction of them
emitted bipolar jets--- much of their diversity could be due to the angle
from which we see them.  Exploiting this possibility to its extreme,
i.e.~using SN~1998bw as an ansatz standard candle, Dar (1999a) suggested at
the `1'st Rome Intl. Workshop on GRBs in the Afterglow Era' that the AGs
of all GRBs may contain a contribution from an SN akin to SN~1998bw, placed
at the GRB's position. However, before GRB~980425 only the redshift of
GRB~971214 (z=3.42) and that of GRB~970508 (z=0.835) were known. An SN~1998bw
displaced to z=3.42 was too faint to be observed, while the data on the
late-time optical AG of GRB~970508 after subtraction of the contribution
from the host galaxy was not accurate enough to allow a conclusive test of
the GRB-SN association.

The first claim of a possible detection of an SN light in the afterglow of
a GRB was made by Bloom et al.~(1999) who interpreted a faint red light
``bump'' superimposed on the late-time afterglow of GRB~980326 as that due
to an SN akin to SN~1998bw.  However, the unknown redshift of GRB~980326 and
the sparse data on its late-time afterglow did not allow a definite
conclusion. Shortly after publishing their paper on the possible evidence
of an SN light in the AG of GRB~980326, four leading authors of this paper
reported on behalf of the Caltech-CARA-NRAO GRB Collaboration the
measurement of the redshift of GRB~970228 (Djorgovski, Kulkarni, Bloom \&
Frail~1999), but, surprisingly, they did not proceed to examine whether
the late-time ``bump'' in the optical light curve of the AG of GRB~970228
is due to an SN akin to SN~1998bw.  This was examined by Dar~(1999b) as soon
as the redshift of GRB~970228 became known, who reported that the addition
of the light of ``a standard candle SN~1998bw at the GRB redshift, z=0.695,
explains better the behavior of the optical afterglow of GRB~970228 than an
extrapolated power-law alone and is consistent with the measured
afterglows of all other GRBs''.  This was immediately confirmed for
GRB~970228 by Reichart~(1999) and later by Galama et al.~(1999). Not only the
magnitude and the peak time of the `bump' were correctly predicted, but
also its broad band spectrum and its evolution coincided with those of
SN~1998bw displaced to the GRB position.

These findings prompted a search for supernova bumps in the afterglow of
other, relatively nearby GRBs (in GRBs with $z>1.2$, the light of a
standard candle SN~1998bw displaced to the GRB position becomes too faint
in the visible bands, in particular, if it suffers also a considerable
extinction by dust in the host and/or Milky Way). However, the AGs of
GRB~990712, GRB~980613 and GRB~980703 did not show clear SN bumps (e.g.,
Hjorth et al.~1999, Sahu et al.~1999; Holland et al.~2000).  But, lack of
evidence for an underlying SN is not evidence against the presence of a
SN.  SNe produce visible bumps in the optical afterglows of nearby GRBs
only if the host galaxy and the jet are not too bright relative to the SN
whose light may also suffer a considerable extinction in the host. 
Moreover, in many cases lack of high quality observations at late time,
lack of a reliable estimate of the late-time AG from its early time
behaviour (when the contributions of the SN and the host galaxy are
negligible) and lack of a reliable method for estimating the extinction of
the SN in the host galaxy, prevented the discovery of an underlying SN.
The last two deficiencies were eliminated once the cannonball (CB) model
of GRBs (Dar \& De R\'ujula 2000; 2003) was demonstrated to provide a good,
simple and universal description of the AGs of all the GRBs of known
redshift (Dado et al. 2002a; 2003a). In the CB model GRBs are produced by
highly relativistic jets of plasmoids (cannonballs) of ordinary matter
ejected in SN explosions akin to SN~1998bw. In the CB model analysis of
these AGs, significant evidence or clear hints were found for a GRB-SN
association in all the cases in which the SN could be observed in practice 
(about 1/2 of the total) and was not found when it could not be seen. This
led Dado, Dar \& de R\'ujula~(2002a) to conclude that perhaps all long
duration GRBs are produced in SN explosions akin to SN~1998bw (Dar \&
Plaga 1999; Dar \& De R\'ujula 2000; Dado et al. 2002a). Despite the
mounting evidence for a GRB-SN association (GRB~990712: Bjornsson et
al.~2001; Dado et al.~2002a, GRB~980703: Holland et al.~2001; Dado et
al.~2002a, GRB~000418: Dar \& De R\'ujula 2000a;  Dado, Dar \& De R\'ujula
2000a, GRB~991208: Castro-Tirado et al.~2001;  Dado, Dar \& De R\'ujula
2002a, GRB~970508: Sokolov~2001; Dado et al.~2002a, GRB~000911:Lazzati et
al.~2001; Dado et al.~2002 unpublished), GRB~010921: Dado, Dar \& De
R\'ujula 2002d) the majority of the GRB community remained skeptical and
continued to argue that the GRB-SN association is true only for a rare
type of GRB and not for the classical long GRBs as a a whole
(see e.g. Hurley, Sari, \& Djorgovski~2003, Galama~2003, Waxman~2003).

Scientific theories must be falsifiable. Thus, in order to
test/demonstrate the validity of the CB model and its underlying GRB-SN
association, in four cases of relatively nearby GRBs, Dado, Dar \&
De R\'ujula used the CB model fit to the early-time AG to {\it predict} the
late-time AGs where an SN contribution should become observable.  In all
four cases, GRB~011121 (Dado et al. 2002b), GRB~020405 (Dado et al.
2002c), GRB~0212111 (Dado et al. 2003b) and GRB~030329 (Dado et al. 2003c),
we have successfully {\it predicted} the discovery of an SN akin to
SN~1998bw and what its contribution would be in the different optical
bands. In particular, in the case of GRB~030329 we foretold the day (April
8, 2003) that the underlying SN would be discovered {\it spectroscopically}
(Dado et al.~2003c). In all these cases of relatively nearby GRBs, we
urged the observers to do the obvious: to try to verify spectroscopically
the existence of an underlying SN and its type. Indeed, it was the
dramatic spectroscopic discovery on April 8, 2003 of SN~2003dh, in the very
bright afterglow of the relatively nearby GRB~030329 (Garnavich et al.~2003;
Staneck et al.~2003; Hjorth et al.~2003) which finally convinced many of
the GRB and SN communities that long GRBs are produced in SN explosions
akin to SN~1998bw. 

In the CB model bursts such as GRB~980425 and XRFs are GRBs viewed further
off axis (Dar \& De R\'ujula~2000,~2003; Dado, Dar \& De R\'ujula~20003d.
Their larger viewing angles result in a much smaller fluence or photon
number-count. Consequently, only relatively close by XRFs (i.e. XRFs with
relatively small $z$) are detected with the current sensitivities of the
X-ray satellites. Thus the optical AG of most XRFs should contain a
detectable light from an SN akin to SN~1998bw at the XRF's position which
peaks around 15- 20 days after the XRF. Indeed the optical afterglow of
all well localized XRFs show evidence for an SN peaking around that time
(GRB~980425: Galama et al.~1998; XRF~020903: Soderberg et al~2002;
XRF~030723: Fynbo et al.~2003, 2004; Tominaga et al. 2004; XRF~031203: 
Bersier et al.~2004; Tagliaferri et al.~2004; Thomsen et al.~2004; Gal-Yam
et al.~2004).

\section{The Current Evidence For the GRB/XRF-SN Association}

\subsection{Indirect Evidence}

More than 90\% of core collapse SNe take place in star formation regions. 
If long duration GRBs are produced in SN explosions, then GRBs are 
expected to be located mainly in star formation regions. Indeed:

\begin{itemize}

\item{}
The well localized GRBs were found to be located in star formation region
in distant galaxies (e.g. Paczynski, 1998b; Holland \& Hjorth 1999).

\item{}
The blue colours and/or precise spectroscopy  of the host 
galaxies of GRBs indicate that they are mainly star forming  galaxies.
(Djorgovsky et al.~2001; Bloom, Kulkarni \& Djorgovski~2002).

\item{}
The optical afterglows of GRBs indicate that the GRBs environment
is that expected from SN explosions in star formation regions
(Dado, Dar \& De R\'ujula~2003c, 2004).

\end{itemize} Finally, the remarkable success of the CB model, which is
based on the GRB/XRF-SN association, in providing a good, simple and
universal description of GRBs and XRFs (Dar \& De R\'ujula 2003; Dado et
al. 2003f and references therein) and of their afterglows (Dado et
al.~2002a,b,c;  2003a,b,c,d; 2004 and references therein) has been 
strong evidence for the GRB/XRF-SN association. 

\subsection{CB model evidence from the prompt $\gamma$-ray emission} 

\begin{figure}
\vskip -1cm
\hskip 2truecm
\begin{center}
\epsfig{file=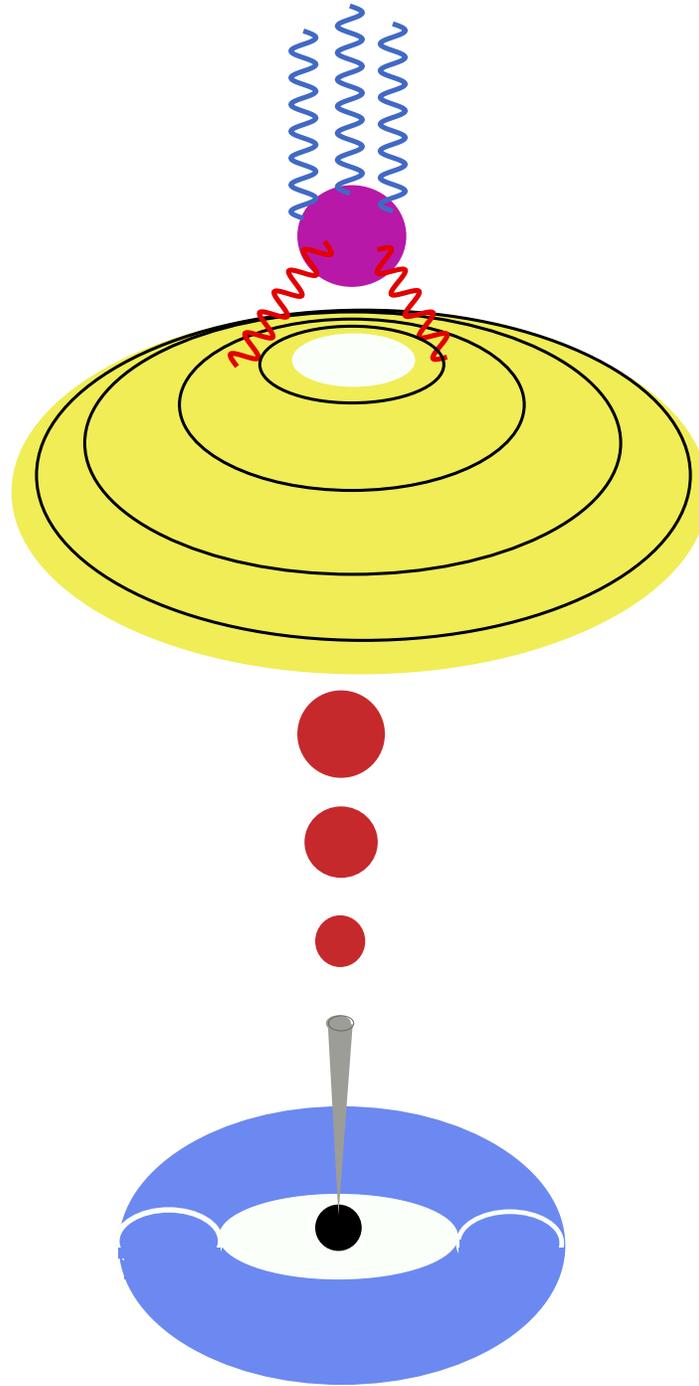, height=20.cm}
\end{center}
\vspace{-.7cm}
\caption{An ``artist's view'' (not to scale) of the CB model
of GRBs and their afterglows. A core-collapse SN results in
a compact object and a fast-rotating torus of non-ejected
fallen-back material. Matter (not shown) abruptly accreted
onto the central object produces
a narrowly collimated jet of CBs, of which only some of
the ``northern'' ones are depicted. As these CBs move through
the ``ambient light'' surrounding the star, they Compton up-scatter
its photons to GRB energies.}
\label{figCB}
\end{figure}

\begin{figure}
\vspace*{-8cm}
\begin{center}
\epsfig{file=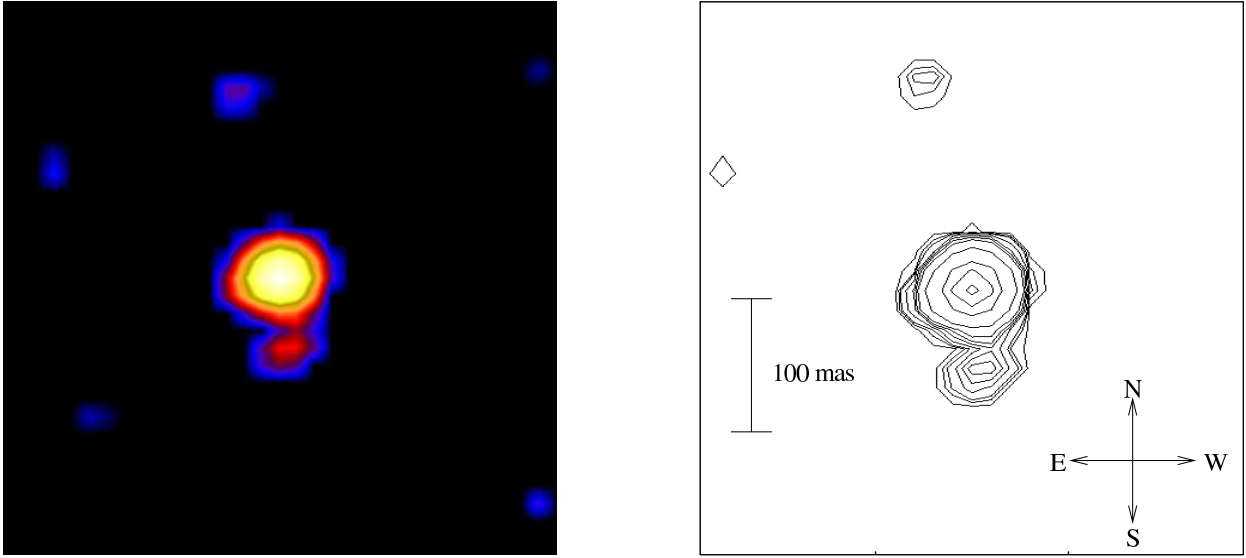, width=16.5cm}
\end{center}
\hskip .5truecm
\caption{The two mystery spots emitted by SN~1987A in opposite axial
directions (Nisenson \& Papaliolios 2000). The northern and southern
bright spots are compatible with being CBs emitted around
the time of the SN explosion and travelling at a velocity equal,
within errors, to $c$. One of the {\it apparent} velocities is superluminal.
The corresponding GRBs were not pointing in our direction.}
\label{figCostas}
\end{figure}

The CB model is illustrated in Fig.(\ref{figCB}). In the CB model,
long-duration GRBs are produced in the explosions of ordinary
core-collapse SNe. Following the collapse of the stellar core into a
neutron star or a black hole, and given the characteristically large
specific angular momentum of stars, it is hypothesized that an accretion
disk or torus is produced around the newly formed compact object, either
by stellar material originally close to the surface of the imploding core
and left behind by the explosion-generating outgoing shock, or by more
distant stellar matter falling back after its passage (De R\'ujula 1987).
A highly relativistic CB is emitted along the rotation axis, as observed
in microquasars, when part of the accretion disk falls abruptly onto the
compact object (e.g.~Mirabel \& Rodrigez 1999; Rodriguez \& Mirabel 1999
and references therein).  Such CBs encounter relatively small baryonic
column densities because lack of rotational support along the rotation
axis results in fast accretion of matter along the polar directions. The
bipolar ejection of relativistic CBs may have been detected in SN~1987A as
shown in Fig.~(\ref{figCostas}) borrowed from Nisenson \& Papaliolios
2000).

The high-energy photons of a single pulse in a GRB/XRF are produced
as a CB coasts through the `glory' surrounding the parent SN. The glory is
the ``echo'' (or {\it ambient}) light from the SN, permeating the
``wind-fed'' circumburst density profile, previously ionized by the early
extreme UV flash accompanying an SN explosion, or by the enhanced UV
emission that precedes it.

\begin{figure}[]
\begin{center}
\vspace{.3cm}
\vbox{\epsfig{file=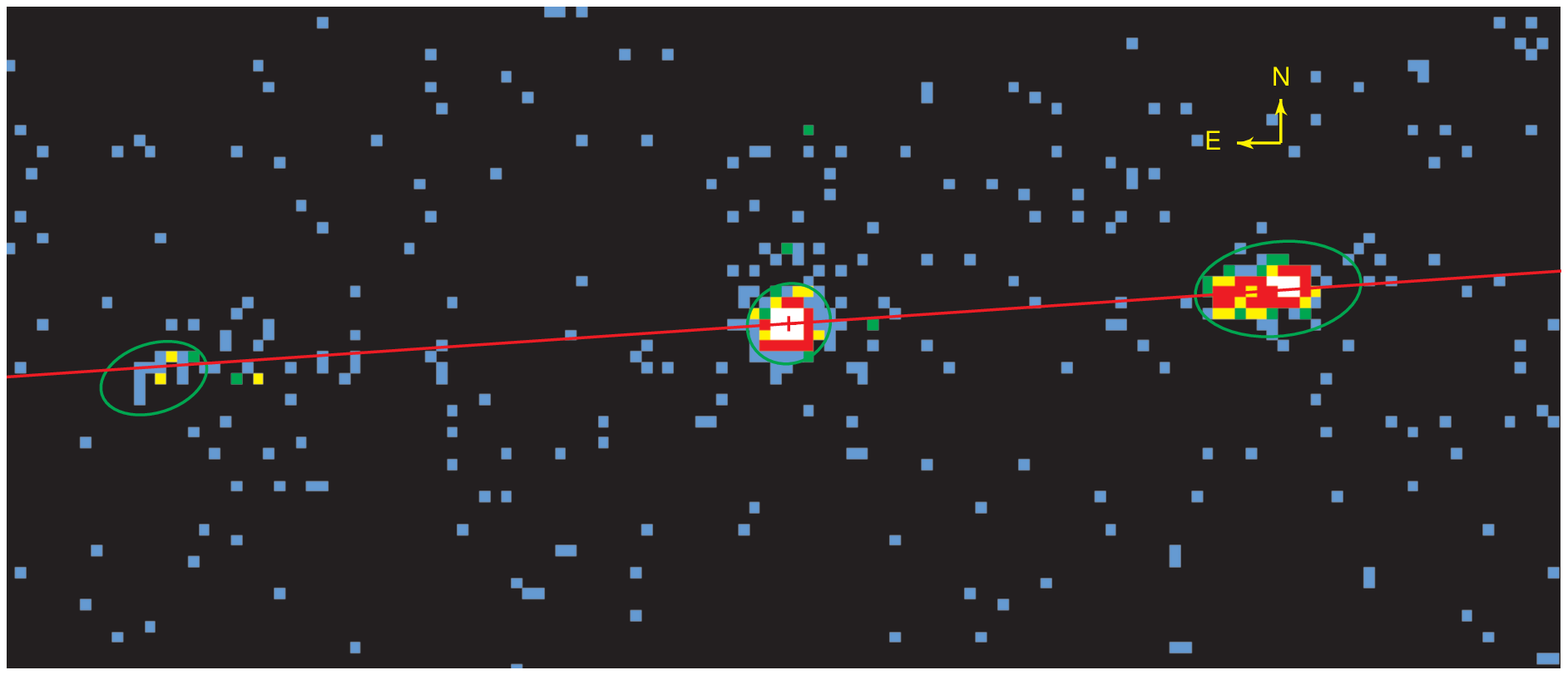,width=13cm}}
\vspace{+1cm}
\vbox{\epsfig{file=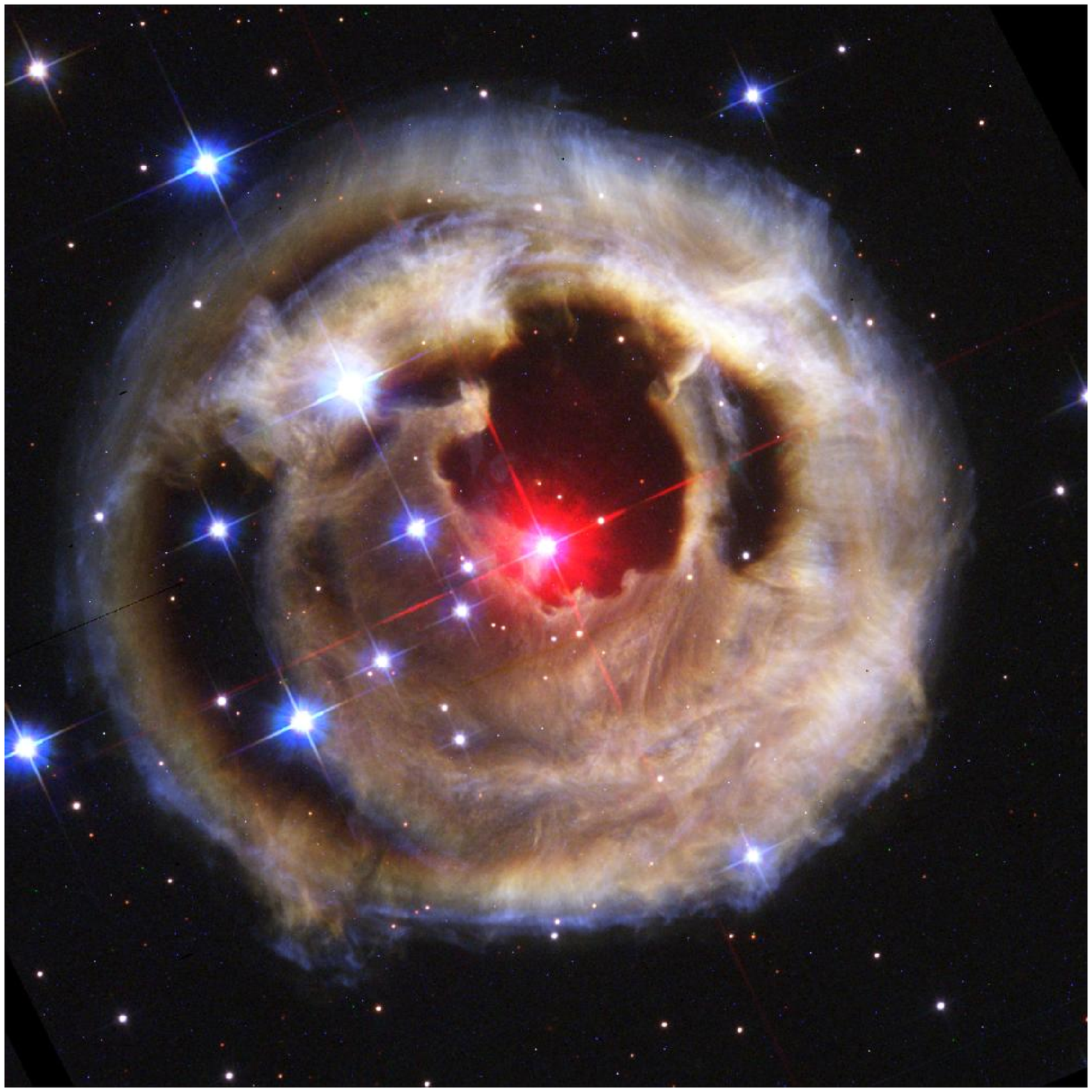, width=13cm}}
%\vspace{-1cm}
\end{center}
\caption{Upper panel: Two relativistic CBs emitted in opposite directions
by the microquasar XTE J1550-564, seen in X-rays by Corbel et
al.~2002. Lower panel:
HST picture from 28 October 2002 of the {\it glory}, or light echo,
of the  stellar outburst of the red supergiant V3838 Monocerosis
in early January 2002. The light echo was formed by scattering off
dust shells from previous ejections (Bond et al.~2003).}
\label{CBGlory}
\end{figure}

The CBs of the CB model are inspired by the ones observed in quasar and
microquasar emissions. One example of the latter is shown in the upper
panel of Fig.~(\ref{CBGlory}), showing two opposite CBs emitted by the
microquasar XTE J1550-564 (Kaaret et al.~2003). The winds and echoes of
GRB-generating SNe are akin to those emitted and illuminated by some very
massive stars. The light echo (or glory) of the stellar outburst of the
red supergiant V3838 Monocerosis in early January 2002 is shown in the
lower panel of Fig.~(\ref{CBGlory}), from Bond et al.~(2003). In a sense
all the CB model results for the prompt gamma-ray emission follow from
superimposing the two halves of Fig.~(\ref{CBGlory}) and working out in
detail what the consequences ---based exclusively on Compton scattering---
are. 

The electrons enclosed in the CB 
Compton up-scatter photons to energies that, close to the CBs direction
of motion, correspond to the $\gamma$-rays of a GRB and
less close to it, to the X-rays of an XRF.
Each pulse of a GRB or an XRF corresponds to one
CB. The timing sequence of emission of the successive individual pulses
(or CBs) reflects the chaotic accretion process and its
properties are not predictable, but those of the single pulses are (Dar \&
De R\'ujula 2003 and references therein). In particular,
the major observed properties of GRB pulses are {\it correctly 
predicted/reproduced} (Dar \& De Rujula 2003). They include: 

\begin{itemize}

\item{}
The characteristic energy $E={\cal{O}}(250)$ keV of the $\gamma$ rays
(Preece et al.~2000; Amati et al.~2002).
\item{}
The narrow distribution of the ``peak'' energies of the GRB
spectra (e.g.~Preece et al.~2000).
\item{}
The duration of the single pulses of GRBs: a median $\Delta t\sim 1/2$ s
full width at half-maximum (McBreen et al.~2002).
\item{}
The characteristic (spherical equivalent) number of photons per pulse,
$N_\gamma\sim 10^{59}$on average, which, combined with the characteristic
$\gamma$ energy,
yields the average total (spherical equivalent) fluence of a GRB
pulse: $\sim 10^{53}$ erg.
\item{}
The general {\it FRED} pulse-shape: a very ``fast rise'' followed by a fast
decay $N(t)\propto 1/t^2$, inaccurately called ``exponential decay"
(Nemiroff et al.~1993, 1994;
Link \& Epstein 1996; McBreen et al.~2002).
\item{}
The $\gamma$-ray energy distribution, $dN/dE\sim E^{-\alpha}$,
with, on average, $\alpha\sim 1$ exponentially evolving into $\alpha\sim 2.1$
and generally well fitted by the ``Band function''
(Band et al.~1993).
\item{}
The time--energy correlation of the pulses: the pulse duration
decreases like $\sim E^{-0.4}$ and peaks earlier the higher the energy
interval
(e.g.~Fenimore et al.~1995; Norris et al.~1996; Ramirez-Ruiz \&
Fenimore 2000; Wu
\& Fenimore 2000); the spectrum gets softer
as time elapses during a pulse (Golenetskii et al.~1983;
Bhat et al.~1994).
\item{}
Large polarizations of the prompt $\gamma$ rays from
GRBs (Coburn \& Boggs 2003) and a much smaller
polarization in XRFs.
\item{}
Various correlations between pairs of the following observables:
photon fluence, energy fluence, peak
intensity and luminosity, photon energy at peak intensity or luminosity,
and pulse duration (e.g.~Mallozzi et al.~1995; Liang \&
Kargatis 1996;
Crider et al.~1999; Lloyd, Petrosian \& Mallozzi~2000; Ramirez-Ruiz \&
Fenimore 2000;
McBreen et al.~2002; Kocevski et al.~2003).
\end{itemize}

\subsection{Direct Evidence in Optical AGs of Nearby GRBs/XRFs}

If GRBs/XRFs are produced by relativistic jets which are
ejected in core collapse SN explosions then 
their AGs consist of three contributions,
from the relativistic jet [RJ], the concomitant SN, and the host 
galaxy [HG] : 
\begin{equation}
F_{AG}=F_{RJ}+F_{SN}+F_{HG}\, ,
\label{sum}
\end{equation}
the latter contribution being usually determined
by late-time observations, when the RJ and SN contributions become
negligible.

Shortly after an SN explosion, the SN brightness increases as a result of
fast expansion of its photosphere. Dissipation of the shock 
energy and decay of the main
radioisotopes, $^{56}Ni$ and $^{56}Co\, $ in the ejecta 
which power the
light emission, reverse this increase and the brightness decays almost 
exponentially. In the case of SN~1998bw the peak brightness was reached 
around $t\approx (1+z)\times 15\, $ days (see e.g., Galama et al.~1998). 
Thus, the contribution of an SN akin to SN~1998bw displaced to the GRB
redshift, is expected to peak around $t\approx (1+z)\times 15$ days.

The colour lightcurves (Galama et al.~1998; McKenzie \& Schaefer 1999; 
Sollerman et al.~2000; Fynbo et al. 2000; Sollerman et al. 2002) and
spectroscopic evolution of SN~1998bw  (Iwamoto et al.~1998; 
Patat et al.~2001; Stathakis et al.~2000) were followed 
in great detail during the first two years after explosion. 
They can be used to estimate the SN contribution,
using SN~1998bw as an ansatz standard candle (Dar 1999b). With this
approximation, 
the spectral energy density of the optical AGs of GRBs with known 
redshift 
$z$ can be written as (Dar \& De R\'ujula 2000; Dado et al.~2002a),
\begin{equation} 
F_{SN}[\nu,t] = 
{1+z \over 1+z_{bw}}\; {D_L^2(z_{bw})\over D_L^2(z)}\, 
F_{bw}[\nu', t']\, A_{SN}(\nu,z,t)\, , 
\label{bw} 
\end{equation} 
where $A_{SN}(\nu,z,t)$ is the attenuation along the line of sight to the SN.
$\nu'=[(1+z)/(1+z_{bw})]\, \nu\, ,$ $t'=[(1+z_{bw})/(1+z)]\, \nu $ and  
$D_L(z)$ is the luminosity distance (we have used a cosmology with 
${\Omega_M}=0.3$, ${\Omega_\Lambda}=0.7$ and $ H_0=65$ km/s/Mpc)
which may depend on time because of sublimation of dust in the host 
galaxy along the line of sight by the GRB and AG.

The SN contribution can be resolved from the optical AG of a GRB and
compared with the `SN~1998bw standard candle' ansatz only if the total AG
and the individual contributions from both the host galaxy and the jet are
known accurately enough. Two main procedures have been employed 
for calculating  $F_{RJ}[\nu,t]$ at late times.  The first procedure,
employed mainly by the observers, used
an ad hoc parametrization (a broken power-law) to extrapolate 
the early time RJ contribution to late times. 
Lacking a theoretical 
basis\footnote{A broken power-law has been frequently used to fit  
optical AGs of GRBs but was never derived theoretically.},
this procedure is unreliable, in particular in cases such as 
GRB~030329 where the AG does not have a simple broken power-law
behaviour. Moreover,
this procedure requires knowledge of the `break time', but
a few attempts to predict these break times from the total GRB fluence 
and the observed early time behaviour have
failed completely.

The second procedure used the CB model fit of the early  
time afterglow in order to calculate $F_{RJ}[\nu, t]$ at late 
time.  In this  model the relativistic jet is made of plasmoids
(cannonballs) of radius $R$ with a bulk motion Lorentz 
factor $\gamma(t)$ moving at an angle $\theta$ relative to the 
line of sight. Their Doppler 
factor $\delta$, for the relevant Lorentz factors and viewing angles,
$\gamma^2 \gg 1$ and $\theta^2 \ll 1$, is given to an excellent 
approximation, by
\begin{equation}
  \delta = {1\over \gamma\, (1-\beta\, \cos\theta)}\approx
           {2\gamma\over 1+\gamma^2\, \theta^2}\; ,
\label{Doppler}
\end{equation}
The deceleration of the CBs in the ISM, i.e. $\gamma(t)\, ,$ 
is determined by energy-momentum conservation
(Dado et al. 2002a).
The AG of a CB is mainly due to synchrotron radiation from
accelerated  electrons in the CB's
chaotic magnetic field. It has the approximate form
(Dado et al.~2003e):
\begin{equation}
F_{RJ}[\nu,t]\propto n^{(1+\hat\alpha)/2}\, R^2\,
\gamma^{3\hat\alpha-1}\,
\delta^{3+\hat\alpha}\, A(\nu,t)\, \nu^{-\hat\alpha}\, ,
\label{afterglow}
\end{equation}
with $\hat\alpha$ changing from $\sim 0.5$ to $\sim 1.1$
as the emitted frequency crosses the
{\it ``injection bend'',}
\begin{equation}
 \nu_b(t) \simeq \nu_0\, \left[{[\gamma(t)]^ 3\, \delta(t) \over 
10^{12}}\right ]\,
\left[{n_p\over 10^{-3}\;cm^3}\right]^{1/2}\, ,
\label{nubend}
\end{equation}
where we estimated (Dado et al.~2003) that $\nu_0\sim 1.87\times 10^{15}$ 
Hz and $n_p$ is 
the baryon density of the interstellar medium.
The attenuation $A(\nu,t)$ is a product of the attenuation in the host
galaxy, in the intergalactic medium, and in our Galaxy. The attenuation in
our galaxy in the direction of the GRB or XRF is usually estimated from the
Galactic maps of selective extinction, $E(B-V)$, of Schlegel, Finkbeiner \&
Davis (1998), using the extinction functions of Cardelli et al. (1989).
The extinction in the host galaxy and the intergalactic medium,
$\rm A(\nu,t)$
in Eq.~(\ref{bw}), can be estimated from the difference between the
observed spectral index {\it at very early time when the CBs are still
near the SN} and that expected in the absence of extinction. Indeed, the
CB model predicts ---and the data confirm with precision--- the gradual
evolution of the effective optical spectral index towards the constant
value $\approx -1.1$ observed in all ``late'' AGs (Dado et al.
2002a, 2003a). The ``late'' index is independent of the attenuation in the
host galaxy, because after a couple of observer days after the 
explosion, the CBs are typically already moving in the low-column-density,
optically-transparent halo of the host galaxy.

Using the CB model, Dado et al.~(2002a,b,c, 2003a,b,c,d,e, 2004) have
shown that the optical AG of {\it all} relatively nearby GRB with known
redshift (all GRB with $\rm z<1.2)$ contains evidence or clear hints for
an SN~1998bw-like contribution to their optical AG, suggesting that most
---and perhaps all--- of the long duration GRBs are associated with
1998bw-like supernovae (in the more distant GRBs, the ansatz standard
candle could not be seen, and it was not seen).  This evidence is
summarized in Table I. The light curves or spectra of AGs which show 
clear photometric and/or spectroscopic evidence for an SN contribution akin
to SN~1998bw are also displayed in Figs. 3-8.  Cases where scarcity of
data, lack of spectral information and multicolour photometry, and
uncertain extinction in the host galaxy prevented a firm conclusion are
listed in the table with a question mark. Similar conclusions were reached
recently by Zeh, Klose \& Hartmann (2003) using a phenomenological
analysis of the late-time AGs of GRBs.

For CBs of constant radius moving in a constant density ISM, 
energy-momentum yields, $\gamma(t)\sim\delta(t)\sim t^{-3}$, and then it 
follows from Eq.~(\ref{afterglow}) that
at late time the RJ contribution decreases like
 $F_{RJ}\sim t^{-2.13}\, \nu^{-1.1}$ 
(Dado et al.~2002a;2003a). For $n_p\sim 1/r^2\, ,$  
energy-momentum conservation implies that at late time, $\delta(t)\sim
2\, \gamma(t)$  approach  a ``constant'' value and then 
Eq.~(\ref{afterglow}) yields  $F_{RJ}\sim t^{-2.10}\, \nu^{-1.1},$
which has practically the same asymptotic behaviour
Thus, in cases where early time AG data is not available, we recommend the
use of this approximate form with a normalization 
fitted directly to the data for subtraction the RJ 
contribution to the late-time AG.

\subsection{GRB~980425/SN~1998bw and the GRB/XRF-SN association}

In the CB model, XRFs are GRBs viewed further off axis 
(Dar \& de R\'ujula 2003, Dado, Dar \& De Rujula 2003f). 
Their properties are similar to those of GRB~980425, which, in the
CB model, was interpreted (Dar \& De R\'ujula 2000, 2003;  Dado et
al.~2002a, 2003a) as an entirely normal GRB produced by the
explosion of SN~1998bw. Its jet of CBs was ejected at
an angle $\theta \sim 3.9/\gamma\sim 8$ mrad, a large value which,
combined with the progenitor's unusually small redshift ($z=0.0085$)
conspired to produce a rather typical GRB fluence (Dar \& De R\'ujula
2000, 2003; Dado et al.~2002a, 2003a). GRB~980425 is by
definition a GRB and not an XRF, as the central value of its peak energy,
$E_p=54.6\pm 20.9$ keV (Yamazaki, Yonetoku \& Nakamura 2003),
is just above the ``official'' borderline 40 keV for XRFs.
In the CB model, SN~1998bw, associated with GRB~980425, is an
ordinary core-collapse SN: its ``peculiar'' X-ray and radio emissions
were not emitted by the SN, but were part of the GRB's AG (Dado
et al. 2002a, 2003a). The high velocity of its ejecta is attributed
to the SN being viewed almost ``on axis".

The larger viewing angles of XRFs result in much smaller Doppler factors,
$\delta$, than those of GRBs.  Relative to GRBs, XRFs have pulses that are
much dimmer in fluence (which is proportional to $\delta^3$) or in photon
number-count (which is proportional to $\delta^2$). Thus, only relatively
close-by XRFs (i.e. XRFs with relatively small $z$) are detected with the
current sensitivities of the X-ray instruments, on board HETE and
Integral. Thus the optical AG of XRFs should contain a detectable
contribution of an SN akin to SN~1998bw, displaced to the XRF's position and
peaking about 20 days after the SN exploded and the XRF (Dar \& De
R\'ujula 2003; Dado et al. 2003f). Such a smoking-gun signature have
already been observed in the AGs of XRF~030723 (Fynbo et al. 2003, 2004) and
XRF~031203: Bersier et al~2004;  Tagliaferri et al.~2004; Thomsen et
al.~2004; Gal-Yam et al.~2004), and are shown in Fig.~(\ref{xrf723}) and
Fig.~(\ref{xrf903}), respectively.

\section{GRB/XRF-SNIa association?}

Little is known for sure about the progenitors or the production
mechanisms of Type Ia SNe. The prevailing theory is that accretion onto a
$C/O$ White Dwarf (WD) from a companion star in a close binary
system causes their collapse ---accompanied by a thermonuclear
explosion--- when the accreting WD's mass exceeds the Chandrasekhar limit
(Whelan \& Iben 1973).  In the case of a WD--WD binary, the trigger may
also be a merger, the end-result of a shrinking of the orbit due to
gravitational-wave emission (Iben \& Tutukov~1984; Webbink~1984).

In every one of the quoted scenarios, the specific angular momentum of the
collapsing system is likely to be large. It is natural to expect that the
collapsing object may have an axial symmetry leading to the bipolar
ejection of jets of CBs, as in quasars, microquasars and the core-collapse
SNe responsible ---in the CB model--- for long-duration GRBs. About
$70\%\pm10\%$ of all SN explosions in the local Universe are of the
core-collapse types, the rest being Type Ia SNe (Tamman, Loeffler \&
Schroder 1994; van den Bergh \& McClure 1994). Intriguingly, $\sim 75\%$
of all GRBs are long and the rest are short. The coincidence may not be
accidental. As discussed in detail in Dar \& De R\'ujula (2003), the
environment of Type-Ia SNe appear to be scaled-down versions of the
corresponding properties of core-collapse SNe. The stage is naturally set
to suspect that short GRBs may be a time-contracted ---but otherwise very
similar--- version of the long GRBs.

The progenitors of core-collapse SNe are short-lived massive stars.
Consequently, most of their explosions take place in star-formation 
regions, in
superbubbles  produced by the winds of
massive stars and the ejecta from previous SNe. The ISM density in these
bubbles is $n\sim 10^{-2}$--$10^{-3}$ cm$^{-3}$.
The progenitors of Type Ia SNe are long-lived and are not confined to
star-formation regions. Their explosions take place in a normal ISM
of typical density  $n\sim 0.1$--$1.0$ cm$^{-3}$.  
For CBs with a baryon number 100 times smaller in short
GRBs than in long ones (Dar \& De R\'ujula 2003), and even for a density 
around a short GRB
as low as $n=10^{-2}$ cm$^{-3}$, the characteristic time of decline of
the AGs of short GRBs is $\sim 50$ times shorter than for long ones.
Moreover, the smaller CB's radius
($R_\infty^2 \sim [N_{CB}/n]^{2/3}$;
Eq.~(16) of
Dado et al.~2002a) also reduces
the intensity of the AGs  considerably.
When the CBs enter the ISM
(within a couple of minutes of observer time), the combination
of these effects makes the AGs of short GRBs
much harder to detect than those of long ones.
The only chance to detect the AG of short GRBs
is at the very early time when the CBs plough through the
short-range circumstellar wind.
Indeed, very early X-ray AGs of short GRBs, declining rapidly with time,
have actually been detected tens to hundreds of seconds after burst
(e.g.~Frederiks et al.~2003).

In long GRBs the AG is a ``background'' that made it difficult for the GRB
community to consider the possibility that they are all associated with
SNe, as they are in the CB model (in which this background is very well
understood). One redeeming feature of the fact that the AGs of short GRBs
decline so fast is that there will be no background to the detection of a
potentially associated Type Ia SN. Moreover, the peak bolometric
luminosity of SNe of type Ia, $ L_{Ia}\approx 10^{43.35}$ erg s$^{-1}$,
reached around $t\sim(1+z)\times 20$ days after burst (e.g.~Leibundgut \&
Suntzeff 2003), is much larger than that of a core-collapse SNe. If these
SNe were to be found in the directional error boxes of short GRBs, they
could be used to localize them, to identify their host galaxies and their
location within them, and to measure their redshifts. This may
significantly increase the detection rate of Type Ia SNe at cosmological
distances.

\section{Concluding Remarks}

The observational data on well localized long GRBs/XRFs clearly indicate
that most, perhaps all, long duration GRBs/XRFs are produced in SN
explosions akin to 1998bw.  The view held by the majority of the GRB
community, that GRB~980425 belongs to a rare class of dim GRB, is losing
ground. Deviations from the ``standard candle ansatz''
for SNe associated with GRBs are expected and may already 
been observed (GRB~0212111: Della Valle et al.~2003; XRF 030723: Fynbo et 
al.~2004; XRF~030903: Cobb et al. 2004; GRB 020104: Levan et al.~2004 and  
GRB~020305: Gorosabel et al.~2004 in preparation) . But, it is premature to
conclude how accurate the standard candle ansatz is. Observed
deviations can be intrinsic, but they may also result from measurement
errors, inaccurate knowledge of the contribution from the host galaxy at
different wavelengths and inaccurate estimates of the intrinsic AG of the GRB
at late time. Even if the deviations are genuine, they may partly result from
different circumburst environments, from time dependent extinction along
the line of sight in the host galaxy due to sublimation of dust by the GRB and
AG, and due to contamination of the SN light by light echoes from the GRB
and the very bright phase of the AG (perhaps also present in the template
SN~1998bw).

So far, the observational data on the GRB/XRF-SN association have revealed
only the tip of an iceberg. In fact, little is known about the GRB-SN
association: We do not know yet whether only a subclass of Type Ib/c SN
explosions (hypernovae: Iwamoto et al.~1998; Woosley~1999) produce GRBs or
whether all SNe of Type Ib/c, or perhaps SNe of Type II as well (Dar \&
Plaga~1999; Dar \& De Rujula~2000), produce GRBs/XRFs. We do not know
whether short duration GRBs/XRFs are also associated with SN explosions,
in particular with SNe Type Ia (Dar \& De R\'ujula~2003). We do not know
whether the progenitors of long GRBs/XRFs are single stars or binary
stars.  We do not know: what are the compact remnants of long GRBs/XRFs --
neutron stars?  strange-quark stars? or stellar black holes? We also do
not know whether these compact remnants are very active for a considerable
time after birth, e.g., as soft gamma ray repeaters, anomalous X-ray
pulsars or microquasars (if the progenitor star was in a close binary). 
Or do they cool quickly into a relatively quiet remnant? 

As in quasars and in microquasars, the production mechanism of highly
relativistic and narrowly collimated jets in SN explosions is unknown. Nor
do we know the exact mechanism which explodes SNe. We even do not known
whether the GRBs are produced promptly in the SN explosions, as advocated
by the collapsar model of GRBs (Woosley et al.~1999), or whether they are 
produced in a
second bang driven by fall-back material onto a protoneutron star and/or
loss of its rotational and thermal support within hours after core
collapse (De R\'ujula~1987) as advocated by the cannonball model of GRBs
(Dar \& De Rujula~2000,2003)\footnote{The observational data on the
GRB-SN association does not seem to support the supranova model (Vietri
and Stella 1998) where GRBs are produced months or years after a core
collapse SN explosion by a delayed collapse of the neutron star to a
strange-quark star or a black hole due to loss of rotational and thermal
support (e.g., Dar 1999b), the supranova model is not ruled out as a
mechanism for short duration GRBs.}.

Finding the correct answers to these questions will require enormous 
efforts, both
observational and theoretical, which, no doubt, will take a long time and
will require a lot of ingenuity and good luck. But some answers are bound
to be provided soon. It is hoped that a large sky coverage and a fast
localization on board of the prompt $\gamma$-ray emission with advanced
$\gamma$-ray telescopes such as SWIFT, and that a rapid multiwave follow up
in the afterglow phase that continues with a high precision until late
times, will provide some of the answers.  Some answers may also come from
gravitational wave detectors like LIGO, high energy $\gamma$-ray
telescopes like GLAST and high energy neutrino telescopes such as ICECUBE and
ANTARES. The Hubble and Spitzer space telescopes together with the present
and future armada of very large ground based telescopes may provide
reliable light curves and perhaps spectra of SNe in AGs of GRBs with $z>1$.
But irrespective of these, the GRB/XRF-SN association with star formation
may provide a powerful tool for studying the history of star formation in
the universe and the expansion rate of the universe -- if GRBs are
produced by nearly `standard candle' SNe as indicated by the
cannonball-model analysis of the afterglows of GRBs with known redshift. 

\noindent {\bf Acknowledgement:} Long term exciting collaboration 
with S. Dado and A. De Ru'jula is gratefully acknowledged.  This research 
was supported by the Asher Space Research Fund at the Technion.

\begin{deluxetable}{llccc}
\tablewidth{0pt}
\tablecaption{Photometric (P) and spectroscopic (S) evidence for SN 
light in GRBs/XRFs with known redshift}
\tablehead{
\colhead{GRB/XRF}  & \colhead{$z$} &\colhead{SN} &
\colhead{$E_{iso}$[erg]} &\colhead{evidence}}
\startdata
980425 & 0.0085& Y   & $7.8\times 10^{47}$ &  P+S      \\
031203 & 0.1055& Y   & $3.0\times 10^{49}$ &  P        \\
030329 & 0.1685& Y   & $1.1\times 10^{52}$ &  P+S      \\
020903 & 0.251 & ?   & $1.1\times 10^{49}$ &  P        \\
011121 & 0.360 & Y   & $4.6\times 10^{52}$ &  P        \\
990712 & 0.433 & Y   & $5.3\times 10^{52}$ &  P        \\
010921 & 0.451 & Y   & $1.4\times 10^{52}$ &  P        \\
020405 & 0.69  & Y   & $7.2\times 10^{52}$ &  P        \\
970228 & 0.695 & Y   & $1.1\times 10^{52}$ &  P        \\
991208 & 0.706 & Y   & $1.5\times 10^{53}$ &  P        \\
970508 & 0.835 & ?   & $1.0\times 10^{52}$ &  P        \\
000210 & 0.846 & ?   & $1.7\times 10^{53}$ &  P        \\
980703 & 0.966 & ?   & $2.3\times 10^{53}$ &  P        \\
021211 & 1.006 & Y   & $6.0\times 10^{51}$ &  P+S      \\
991216 & 1.020 & ?   & $5.4\times 10^{52}$ &  P        \\
000911 & 1.058 & Y   & $4.0\times 10^{53}$ &  P        \\
980613 & 1.097 & ?   & $5.4\times 10^{51}$ &  P        \\
000418 & 1.118 & ?   & $2.6\times 10^{53}$ &  P        \\
020813 & 1.225 & N   & $7.8\times 10^{53}$ &           \\
010222 & 1.477 & N   & $8.6\times 10^{53}$ &           \\
990123 & 1.600 & N   & $2.0\times 10^{54}$ &           \\
990510 & 1.619 & N   & $5.0\times 10^{53}$ &           \\
030226 & 1.989 & N   & $6.5\times 10^{52}$ &           \\
000301c& 2.040 & N   & $4.6\times 10^{52}$ &           \\
000926 & 2.037 & N   & $2.6\times 10^{53}$ &           \\
011211 & 2.141 & N   & $6.7\times 10^{53}$ &           \\
021004 & 2.330 & N   & $5.6\times 10^{52}$ &           \\
971214 & 3.418 & N   & $2.1\times 10^{53}$ &           \\
000131 & 4.500 & N   & $1.2\times 10^{54}$ &           \\
\enddata 
\end{deluxetable}

\clearpage

\clearpage

\clearpage

\clearpage

\begin{figure}[t]
\hskip 2truecm
\vspace*{2cm}
\hspace*{-1.7cm}
\epsfig{file=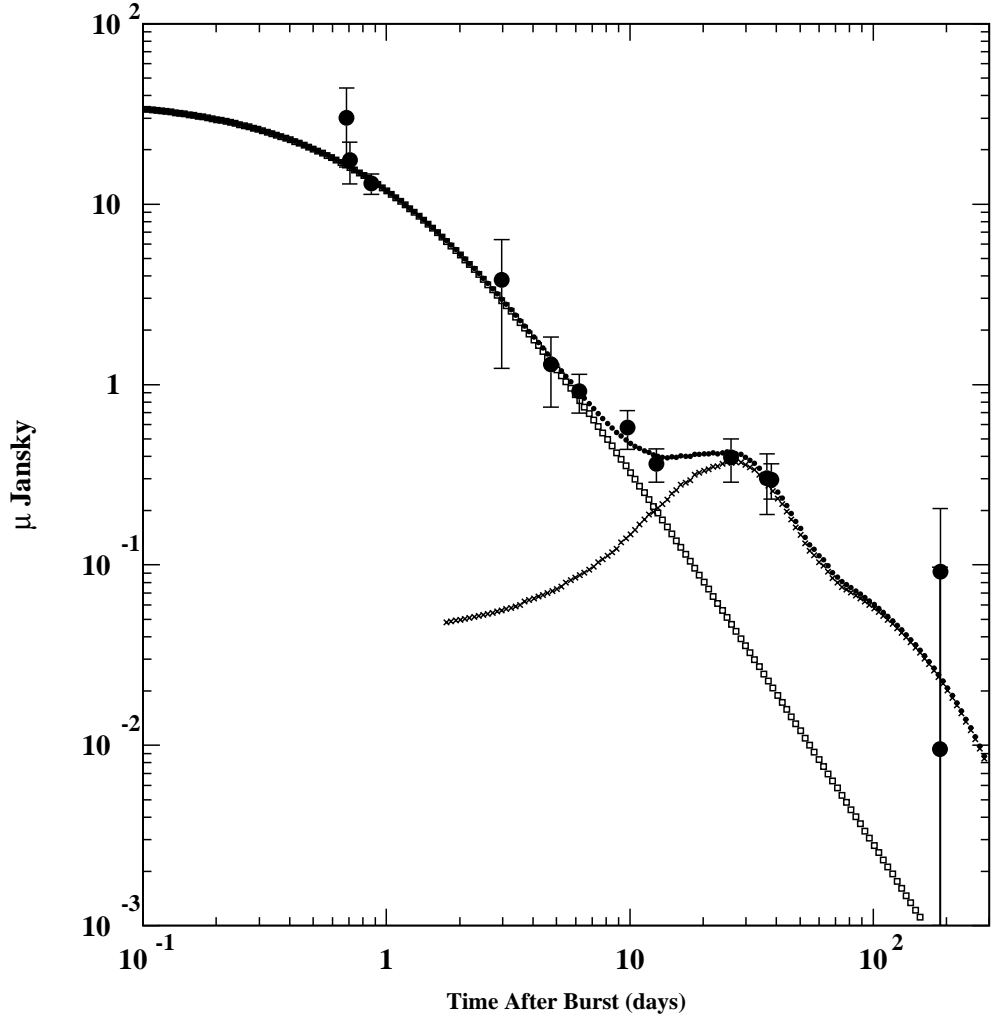, width=15cm} \\
\vspace{-0.5cm}
\caption{The CB model fit (Dado et al.~2002a)
for the R-band AG of GR~970228 [ $z=0.695$],
after subtraction of the contribution from the host galaxy. 
The contribution from
a 1998bw-like supernova placed at the GRB's redshift and corrected for
extinction, is indicated by a line of crosses. The SN bump is clearly
discernible.}
\label{fig228}

\end{figure}

\begin{figure}[t]
\begin{tabular}{cc}
\hskip 2truecm
\vspace*{2cm}
\hspace*{-1.7cm}
\epsfig{file=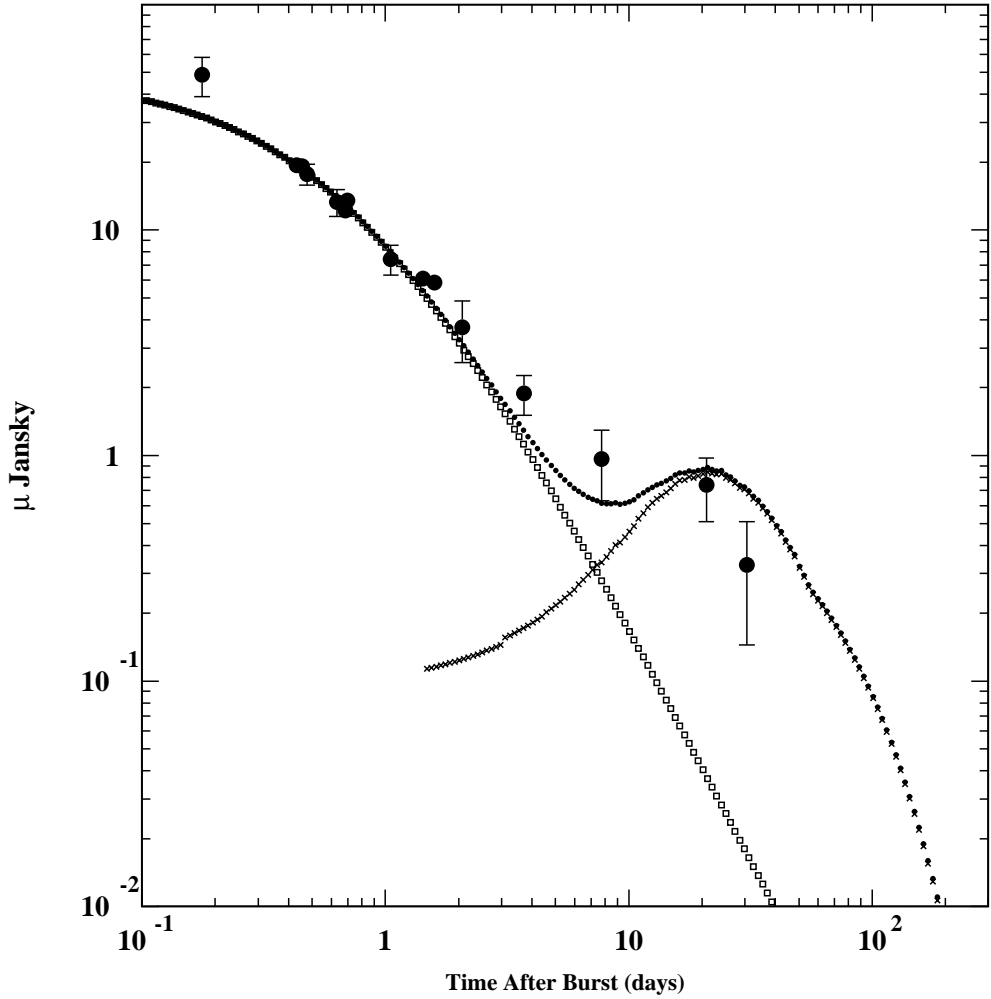, width=15cm}
\end{tabular}
\caption{CB model  fit  (Dado et al.~2002a) to the R band AG of GRB~990712 
[$z=0.433$] after subtraction of the contribution from the host galaxy.
The theoretical contribution  from an SN~1998bw-like supernova
was dimmed by the known extinction in the Galaxy and our
estimated extinction in the host from the early time AG.
The SN  is clearly discernible, but a bump at slightly
earlier times than that of a standard-candle
SN~1998bw would provide a better description.}
\label{fig712}
\end{figure}
%\clearpage

%%%%%%%%%%%%%%%%%%%%%%%%%%%%%%%%%%%%%%

\begin{figure}[t]
\hskip 2truecm
\vspace*{2cm}
\hspace*{-1.7cm}
\epsfig{file=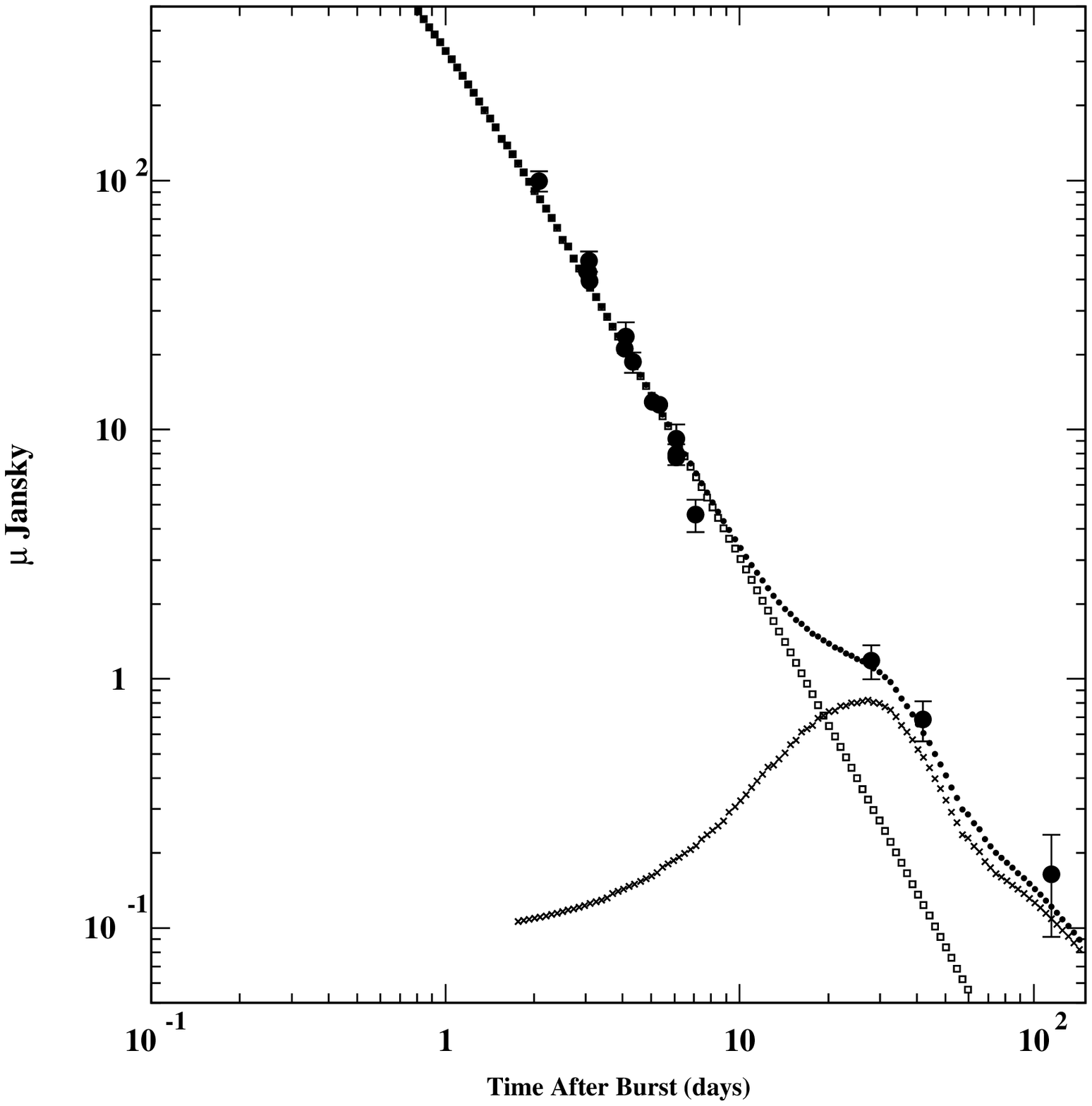, width=15cm} \\
\caption{The CB model fit (Dado et al.~2002a) for
the R-band afterglow
of GRB~991208 [$\rm z=0.706$],
after subtraction of the host
galaxy's contribution.
The contribution
from a 1998bw-like supernova placed at the GRB's
redshift, corrected for  extinction,
is indicated by a line of crosses.
The SN contribution is clearly discernible.}
\label{fig208}
\end{figure}
\clearpage
%%%%%%%%%%%%%%%%%%%%%%%%%%%%%%%%%%%%%%

\begin{figure}[t]
\hskip 2truecm
\vspace*{2cm}
\hspace*{-1.7cm}
\epsfig{file=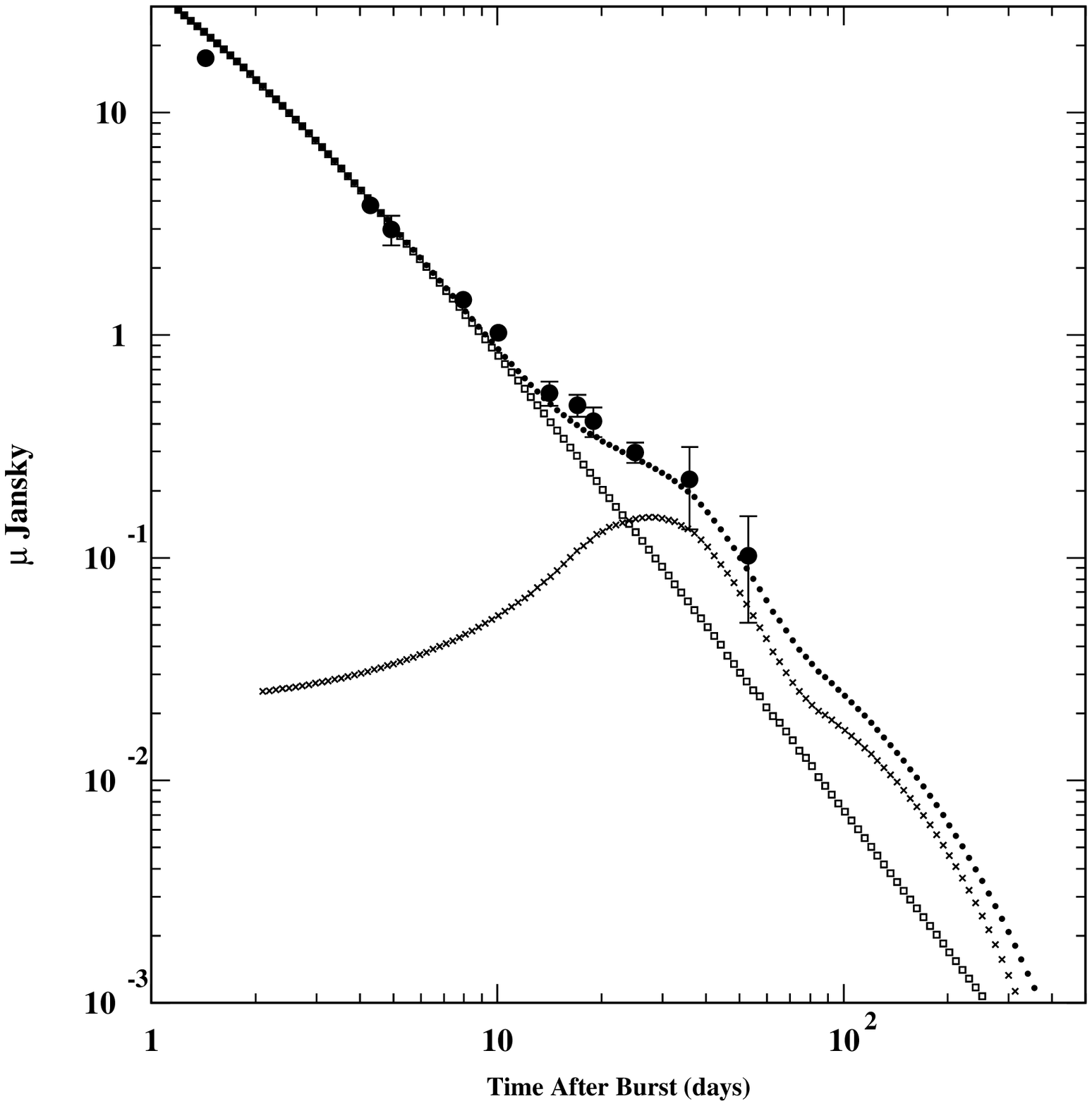, width=15cm} \\
\caption{The CB model fit for the R-band afterglow
of GRB~000911 [$\rm z=1.006$], after subtraction of the host
galaxy's contribution.
The contribution
from a 1998bw-like supernova placed at the GRB's
redshift, corrected for  extinction,
is indicated by a line of crosses.
The SN contribution is clearly discernible.}
\label{fig0911}
\end{figure}
\clearpage
%%%%%%%%%%%%%%%%%%%%%%%%%%%%%%%%%%%%%%
\begin{figure}[t]
\hskip 2truecm
\vspace*{2cm}
\hspace*{-1.7cm}
\epsfig{file=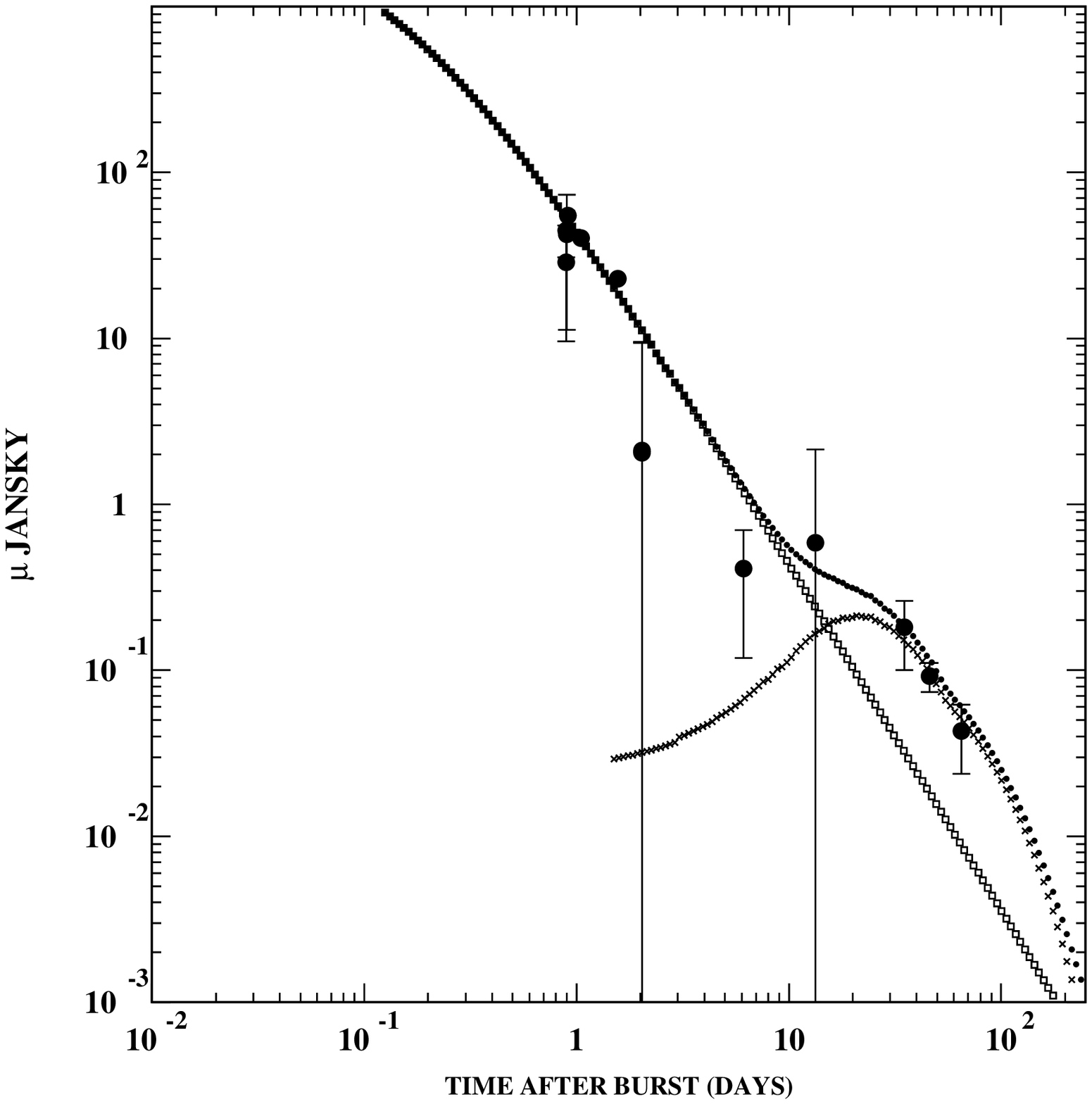, width=15cm} \\
\caption{The CB model fit for the R-band afterglow
of GRB~010921, [$\rm z=0.451$] after subtraction of the host
galaxy's contribution.
The contribution
from a 1998bw-like supernova placed at the GRB's
redshift, corrected for  extinction,
is indicated by a line of crosses.
The SN contribution is clearly discernible.}
\label{fig0921}
\end{figure}
\clearpage
%%%%%%%%%%%%%%%%%%%%%%%%%%%%%%%%%%%%%%

\begin{figure}[]
\hskip 2truecm
\vspace*{2cm}
\hspace*{-2.6cm}
\epsfig{file=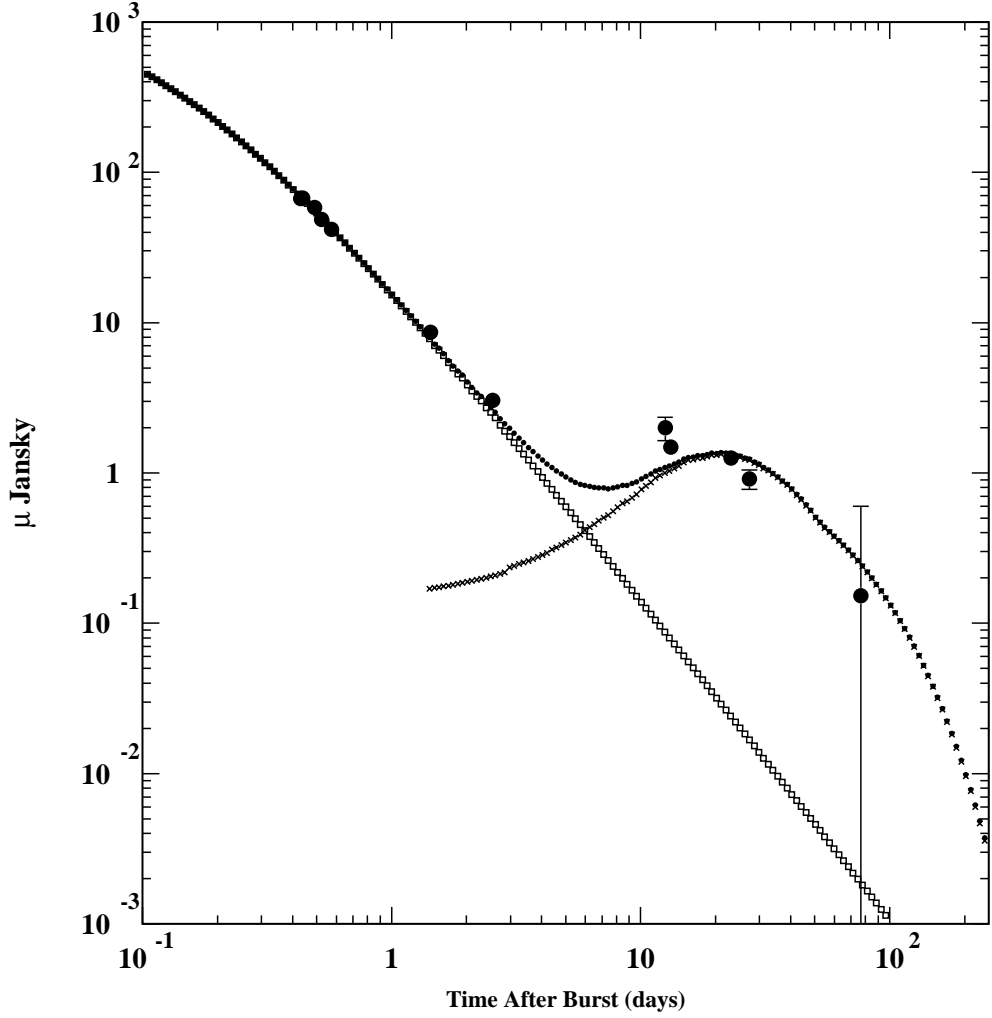, width=15cm}
\caption{CB model  fit (Dado et al. 2002b) for the R-band afterglow
of GRB~011121 [$ z=0.36$] after subtraction of 
the contribution of the host galaxy.
The  contribution  from a 1998bw-like supernova placed at the
GRB's redshift, indicated by a line of crosses,
is corrected by estimated extinction factor.
The SN contribution is clearly discernible.}
\label{figr1121}
\end{figure}

\clearpage

\begin{figure}[t]
%\begin{tabular}{cc}
\vspace{-.5cm}
\hskip 2truecm
%\vspace*{2cm}
\hspace*{-2.1cm}
\epsfig{file=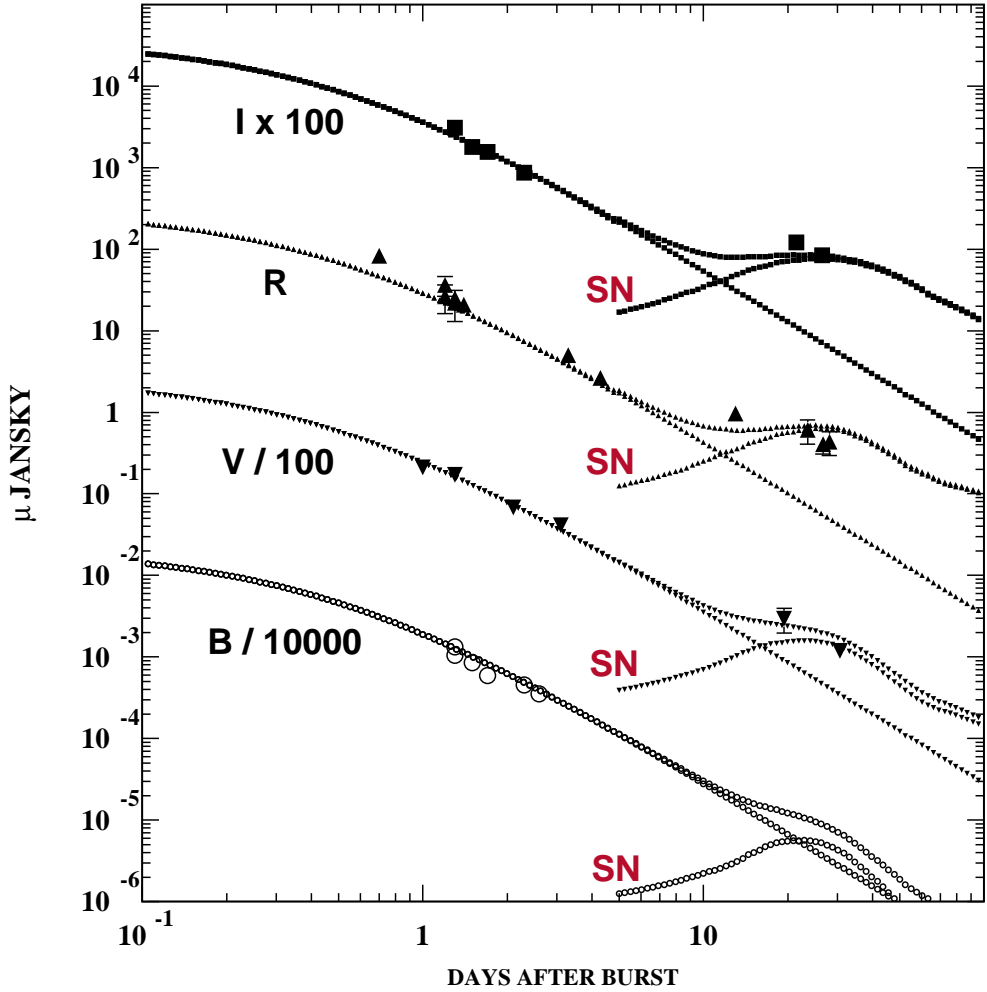, width=15.cm} \\
%\end{tabular}
\vspace{-0.5cm}
\caption{CB model  fit  (Dado et al.~2002c)
to the measured I, R, V, and B-band AG of GRB~020405 [$z=0.69$],
multiplied by 100, 1, 1/100, 1/10000, respectively with the 
the host galaxy contribution subtracted.
The theoretical contribution  from an SN~1998bw-like supernova
was dimmed by the known extinction in the Galaxy and our 
estimated extinction in the host galaxy from the early-time AG. }
\label{figr0425}
\end{figure}
\clearpage

%%%%%%%%%%%%%%%%%%%%%%%%%%%%%%%%%%%%%%%%%%%%%%%%%%%%%%%%

\begin{figure}[t]
%\begin{tabular}{cc}
\vspace{-.5cm}
\hskip 2truecm
%\vspace*{2cm}
\hspace*{-2.1cm}
\epsfig{file=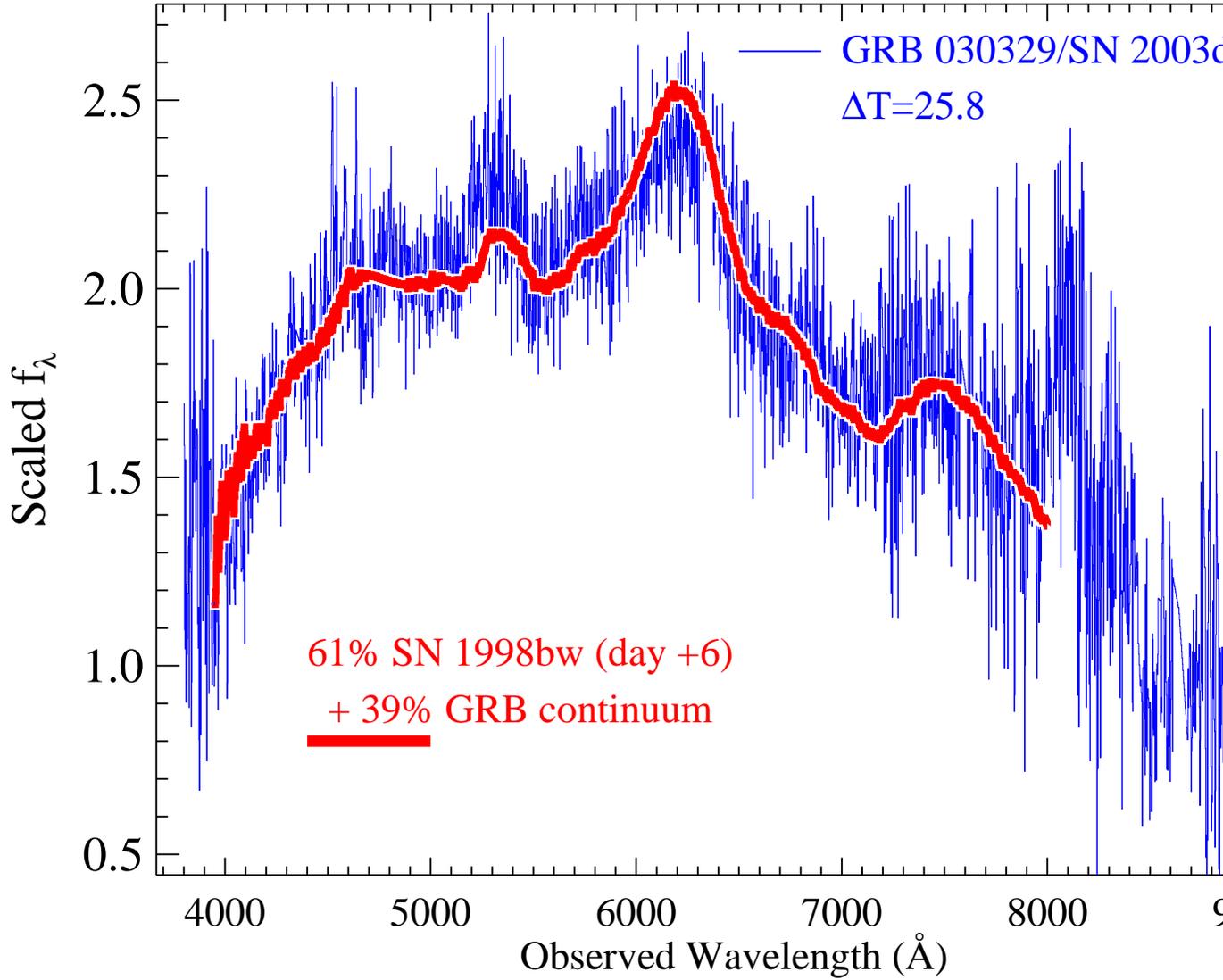, width=16cm, angle=90} \\
%\end{tabular}
\vspace{-0.5cm}
% \begin{figure}
% \begin{center}
%\leavevmode\epsfxsize=10cm \epsfbox{fig4.eps}
%\end{center}
%Figure caption
\caption{Observed spectrum (thin line) of the
GRB~030329/SN~2003dh [$z=0.1685$] afterglow at t=25.8 days after burst.  
The model spectrum (thick line) consists of 39\% continuum and 61\% 
SN~1998bw from 6 days after maximum. The figure is borrowed from 
Matheson et al.~2003}
\label{SN2003dh}
\end{figure}

\clearpage
%%%%%%%%%%%%%%%%%%%%%%%%%%%%%%%%%%%%%%%%%%%%%%%%%%%%%%%%

\begin{figure}
\epsfig{file=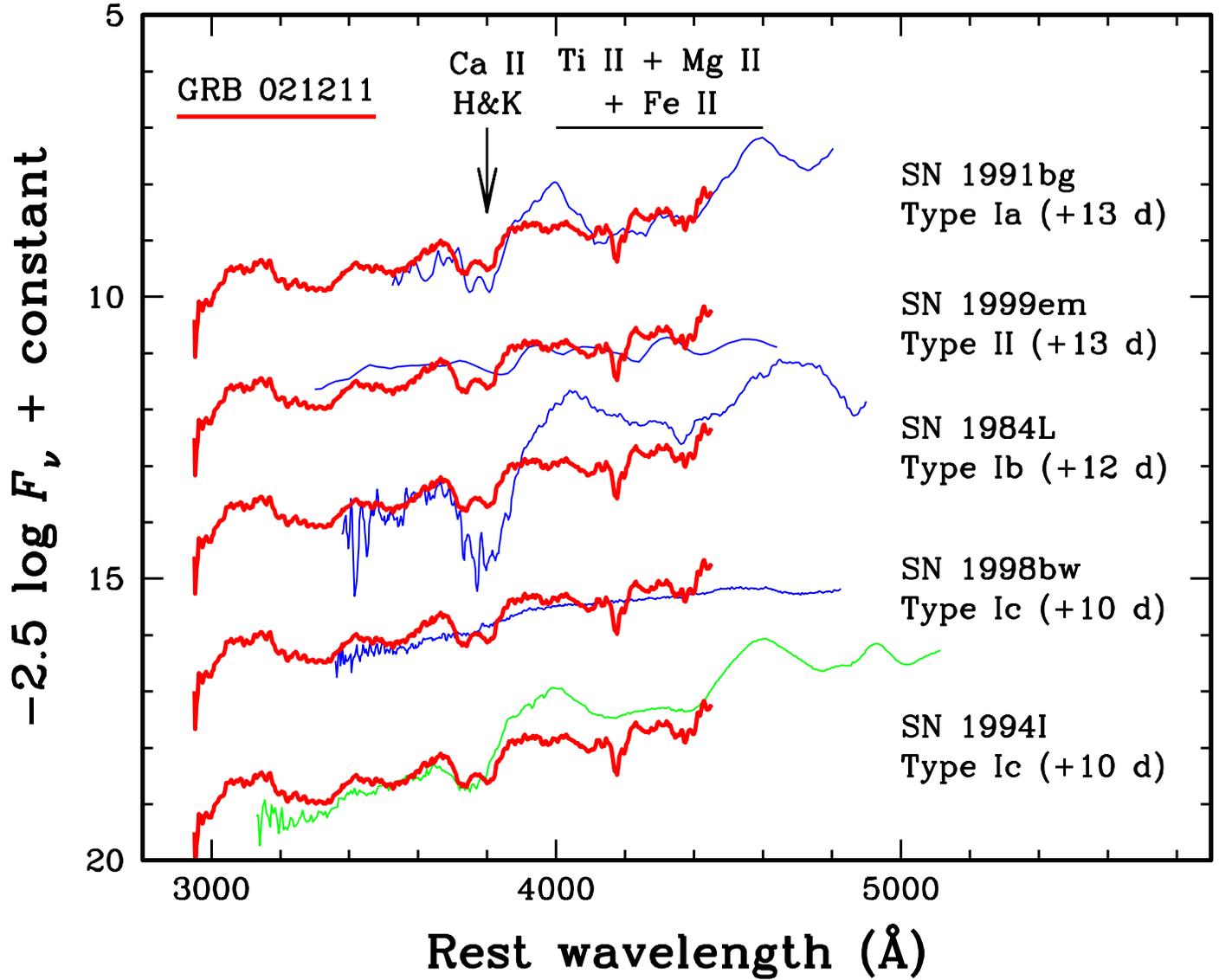, height=15cm}
\vskip 0.0truecm
\caption{Rest-frame spectrum of the afterglow of GRB~021211 [$z=1.006$] 
27~days after the GRB (thick lines), compared
with that of several SNe (thin lines). The figure was borrowed 
from Dela Valle et al.~2003}
\label{fig021211}
\end{figure}

\clearpage
       
%%%%%%%%%%%%%%%%%%%%%%%%%%%%%%%%%%%%%%%%%%%%%%%%%%%%%%%%
\begin{figure}[k]
\hskip 2truecm
\vspace*{2cm}
\hspace*{-2.6cm}
\epsfig{file=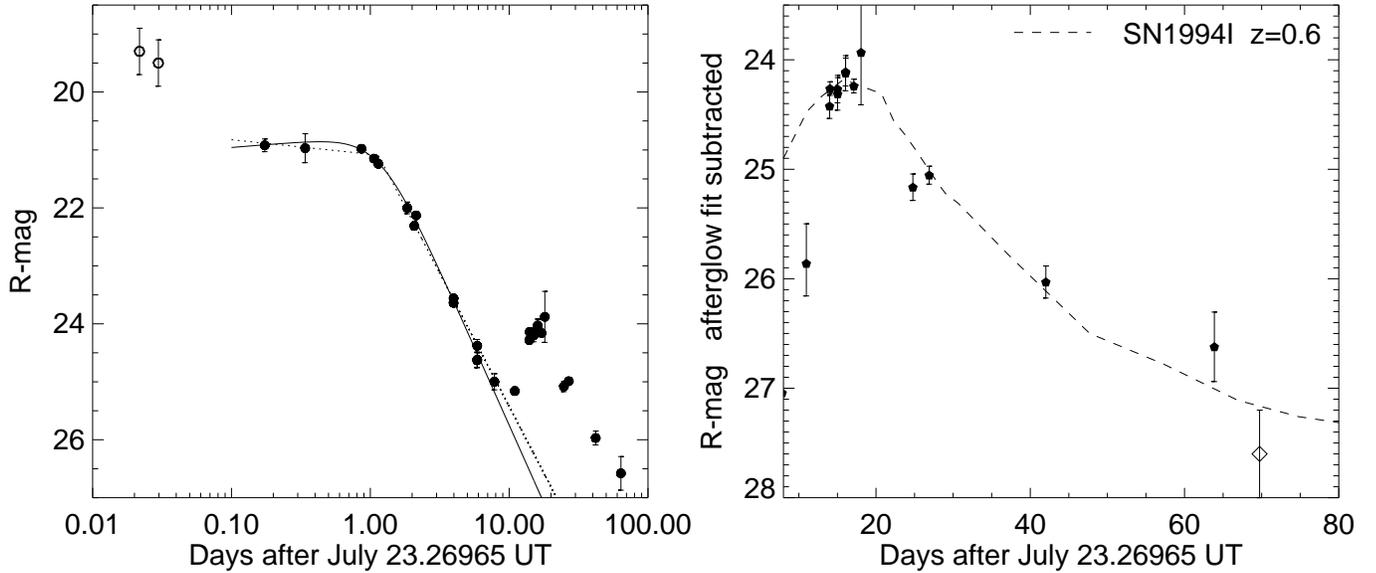, width=18cm}
\caption{ Left: The R-band light curve of the afterglow of XRF~030723
borrowed from Fynbo et al.~2004.
The dashed and solid lines are broken power-law
fits to the  data points. Right: The late time R-band light curve after 
subtraction of the power-law fit. The 
dashed curve shows the B-band light curve of the type Ic SN~1994I
redshifted to $z=0.6$ by Fynbo et al.~(2004) and scaled up in flux by one 
magnitude.} 
\label{xrf723} 
\end{figure}

\clearpage

\begin{figure}[k]
\hskip 2truecm
\vspace*{2cm}
\hspace*{-2.6cm}
\epsfig{file=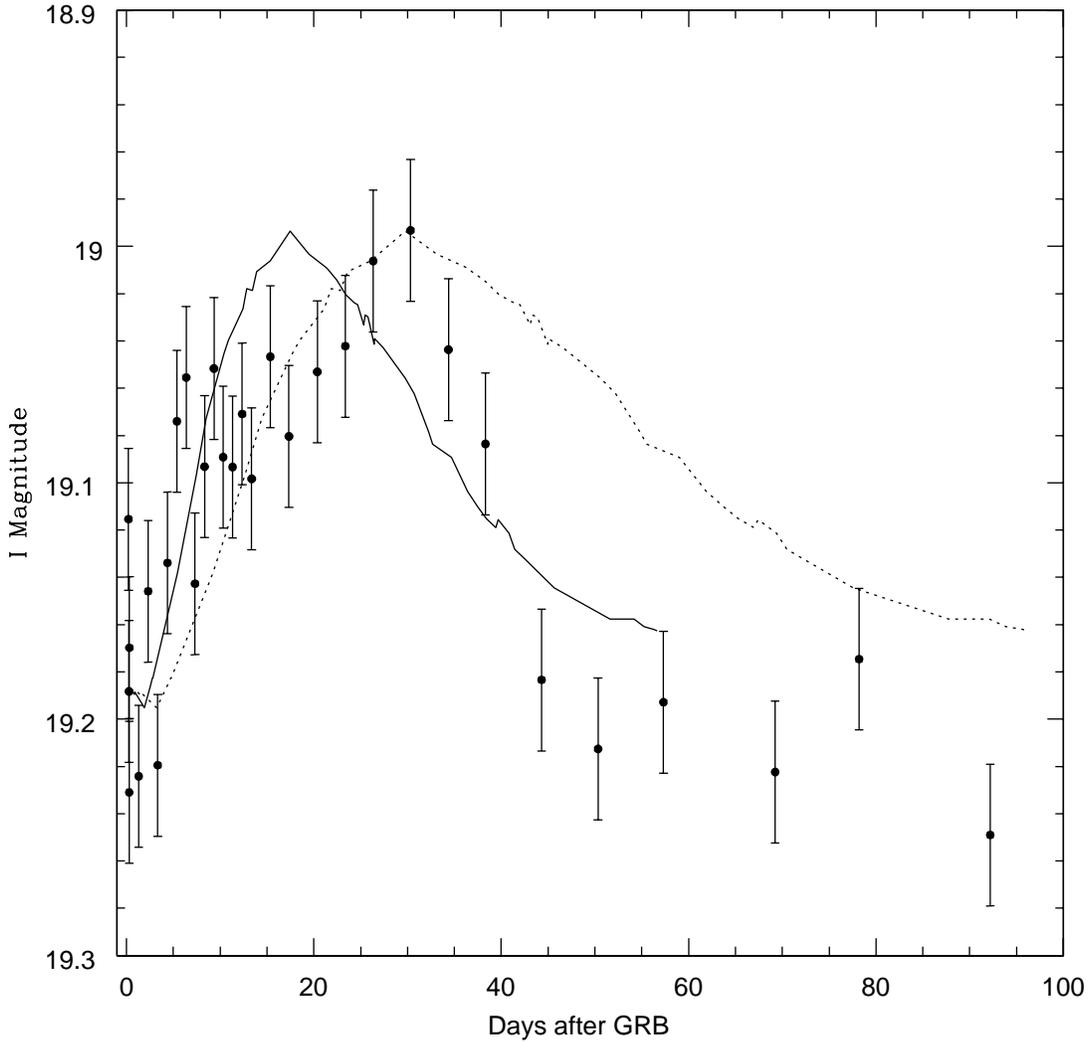, width=15cm}
\caption{The I band lightcurve of the AG of XRF~030903 [$z=0.1055$], 
obtained by  Cobb et al.~2004 by subtracting the host galaxy contribution.
The solid line is a template  SN 1998bw that was displaced to
the GRB position in the host galaxy. 
An additional 0.02 mag has been added to the shifted 
SN 1998bw lightcurve so that the SNe reaches the same peak brightness. If a
stretch of 1.7 is applied (dotted line) the peaks coincide but the declines
is inconsistent with the data.}
\label{xrf903}
\end{figure}

\clearpage

\end{document}